\documentclass[aps,prl,preprint,tightenlines,superscriptaddress,showpacs,byrevtex]{revtex4}
\usepackage{graphicx}
\usepackage{dcolumn}
\usepackage{bm}
\usepackage{pstricks}
\usepackage{pst-node}

\newcommand{\svalue}{-0.67}
\newcommand{\sstaterr}{\pm0.16}
\newcommand{\ssysterr}{\pm0.06}
\newcommand{\avalue}{+0.56}
\newcommand{\astaterr}{\pm0.12}
\newcommand{\asysterr}{\pm0.06}
\newcommand{\ascorrelation}{+0.09}

\usepackage{epsfig}

\usepackage{relsize}
\def\babar{\mbox{\slshape B\kern-0.1em{\smaller A}\kern-0.1em
    B\kern-0.1em{\smaller A\kern-0.2em R}}}

\begin{document}

\resizebox{!}{3cm}{\includegraphics{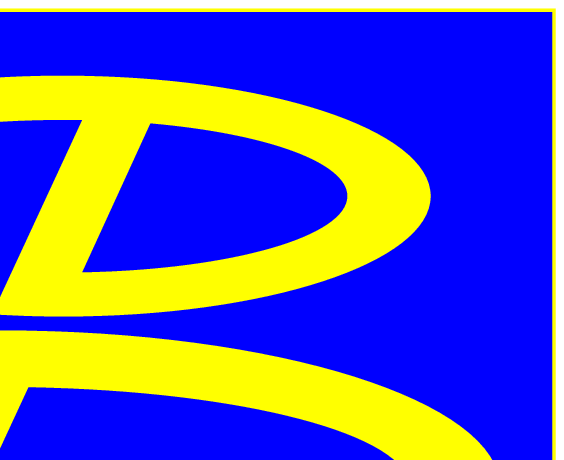}}
\preprint{\vbox{
                 \hbox{KEK    Preprint 2005-9}
                 \hbox{Belle  Preprint 2005-16}
}}
\vspace*{0.1cm}


\title{\quad\\[0.5cm] \boldmath 
{Improved Evidence for Direct {\boldmath $CP$} Violation in $B^0 \to \pi^+\pi^-$ Decays
and Model-Independent Constraints on $\phi_2$}
}

\date{\today}

\affiliation{Budker Institute of Nuclear Physics, Novosibirsk}
\affiliation{Chiba University, Chiba}
\affiliation{Chonnam National University, Kwangju}
\affiliation{University of Cincinnati, Cincinnati, Ohio 45221}
\affiliation{Gyeongsang National University, Chinju}
\affiliation{University of Hawaii, Honolulu, Hawaii 96822}
\affiliation{High Energy Accelerator Research Organization (KEK), Tsukuba}
\affiliation{Hiroshima Institute of Technology, Hiroshima}
\affiliation{Institute of High Energy Physics, Chinese Academy of Sciences, Beijing}
\affiliation{Institute of High Energy Physics, Vienna}
\affiliation{Institute for Theoretical and Experimental Physics, Moscow}
\affiliation{J. Stefan Institute, Ljubljana}
\affiliation{Kanagawa University, Yokohama}
\affiliation{Korea University, Seoul}
\affiliation{Kyungpook National University, Taegu}
\affiliation{Swiss Federal Institute of Technology of Lausanne, EPFL, Lausanne}
\affiliation{University of Ljubljana, Ljubljana}
\affiliation{University of Maribor, Maribor}
\affiliation{University of Melbourne, Victoria}
\affiliation{Nagoya University, Nagoya}
\affiliation{Nara Women's University, Nara}
\affiliation{National Central University, Chung-li}
\affiliation{National United University, Miao Li}
\affiliation{Department of Physics, National Taiwan University, Taipei}
\affiliation{H. Niewodniczanski Institute of Nuclear Physics, Krakow}
\affiliation{Nippon Dental University, Niigata}
\affiliation{Niigata University, Niigata}
\affiliation{Nova Gorica Polytechnic, Nova Gorica}
\affiliation{Osaka City University, Osaka}
\affiliation{Osaka University, Osaka}
\affiliation{Panjab University, Chandigarh}
\affiliation{Peking University, Beijing}
\affiliation{Princeton University, Princeton, New Jersey 08544}
\affiliation{Saga University, Saga}
\affiliation{University of Science and Technology of China, Hefei}
\affiliation{Seoul National University, Seoul}
\affiliation{Sungkyunkwan University, Suwon}
\affiliation{University of Sydney, Sydney NSW}
\affiliation{Tata Institute of Fundamental Research, Bombay}
\affiliation{Toho University, Funabashi}
\affiliation{Tohoku Gakuin University, Tagajo}
\affiliation{Tohoku University, Sendai}
\affiliation{Department of Physics, University of Tokyo, Tokyo}
\affiliation{Tokyo Institute of Technology, Tokyo}
\affiliation{Tokyo Metropolitan University, Tokyo}
\affiliation{Tokyo University of Agriculture and Technology, Tokyo}
\affiliation{University of Tsukuba, Tsukuba}
\affiliation{Virginia Polytechnic Institute and State University, Blacksburg, Virginia 24061}
\affiliation{Yonsei University, Seoul}
   \author{K.~Abe}\affiliation{High Energy Accelerator Research Organization (KEK), Tsukuba} 
   \author{I.~Adachi}\affiliation{High Energy Accelerator Research Organization (KEK), Tsukuba} 
   \author{H.~Aihara}\affiliation{Department of Physics, University of Tokyo, Tokyo} 
   \author{K.~Arinstein}\affiliation{Budker Institute of Nuclear Physics, Novosibirsk} 
   \author{Y.~Asano}\affiliation{University of Tsukuba, Tsukuba} 
   \author{T.~Aushev}\affiliation{Institute for Theoretical and Experimental Physics, Moscow} 
   \author{A.~M.~Bakich}\affiliation{University of Sydney, Sydney NSW} 
   \author{Y.~Ban}\affiliation{Peking University, Beijing} 
   \author{E.~Barberio}\affiliation{University of Melbourne, Victoria} 
   \author{M.~Barbero}\affiliation{University of Hawaii, Honolulu, Hawaii 96822} 
% ...
\collaboration{The Belle Collaboration}

\begin{abstract}
We present a new measurement of the 
time-dependent $CP$-violating parameters
in $B^0\to \pi^+\pi^-$ decays with
$275\times10^6$~$B\overline{B}$ pairs collected 
with the Belle detector at the KEKB asymmetric-energy $e^+e^-$
collider operating at the $\Upsilon(4S)$ resonance.
We find $666\pm43$ 
$B^0\to\pi^+\pi^-$ events and 
measure the $CP$-violating parameters:
${\cal S}_{\pi\pi}  = \svalue\sstaterr ({\rm stat})
                              \ssysterr({\rm syst})$ and
${\cal A}_{\pi\pi} = \avalue\astaterr ({\rm stat})
                              \asysterr ({\rm syst})$. 
We find evidence for large direct $CP$-violation with a significance 
greater than 4 standard deviations for any ${\cal S}_{\pi\pi}$
value.
Using isospin relations, we obtain 95.4\% confidence intervals
for the CKM angle $\phi_2$ of $0^{\circ}<\phi_2<19^{\circ}$
and $71^{\circ}<\phi_2<180^{\circ}$.

\end{abstract}


\maketitle


One of the unresolved mysteries in particle physics is the origin of
$CP$ violation.
Kobayashi and Maskawa (KM) pointed out in 1973 that
$CP$ violation can be incorporated as an irreducible complex phase 
in the weak-interaction quark mixing matrix in the standard model (SM)
framework~\cite{ref:km}.
The KM model predicts $CP$-violating asymmetries in the time-dependent
rates of neutral $B$ meson decays to the $CP$ 
eigenstate $\pi^+\pi^-$~\cite{ref:tcpv}.
In the decay chain of $\Upsilon(4S) \to B^0\overline{B}{}^0 
\to (\pi^+\pi^-)(f_{\rm tag})$, one of the neutral $B$ mesons decays
into $\pi^+\pi^-$ at time $t_{\pi\pi}$
and the other decays at time $t_{\rm tag}$ to a final state $f_{\rm tag}$ 
that distinguishes its flavor.
The time-dependent decay rate is given by
\begin{eqnarray}
%
%
{\cal P}_{\pi\pi}^q(\Delta t) 
& = & \frac{e^{-|\Delta t|/\tau_{B^0}}}{4\tau_{B^0}}
[1 + q \cdot \{ {\cal S}_{\pi\pi} \sin(\Delta m_d \Delta t) \nonumber \\
 & &         + {\cal A}_{\pi\pi} \cos(\Delta m_d \Delta t ) \} ], 
%
%
\label{eq:signal_pdf}
\end{eqnarray}
where $\Delta t = t_{\pi\pi} - t_{\rm tag}$, $\tau_{B^0}$ is the
$B^0$ lifetime, $\Delta m_d$ is the mass difference between
the two neutral $B$ mass eigenstates, and 
$q=+1$ $(-1)$ for $f_{\rm tag} = B^0$ $(\overline{B}{}^0$).
We measure ${\cal S}_{\pi\pi}$ and ${\cal A}_{\pi\pi}$,
which are the mixing-induced and 
direct $CP$-violating parameters, respectively.
In the case where only a $b\to u$ tree transition contributes to the decay
$B^0\to \pi^+\pi^-$~\cite{ref:charge-conjugate}, we would have
${\cal S}_{\pi\pi} = \sin2\phi_2$ and 
${\cal A}_{\pi\pi} = 0$.
Because of possible contributions from $b\to d$ 
penguin transitions that have different
weak and strong phases, 
${\cal S}_{\pi\pi}$ may deviate from $\sin2\phi_2$,
and direct $CP$ violation, ${\cal A}_{\pi\pi} \neq 0$, may occur.
Our previous measurement 
based on a $140$~fb$^{-1}$ data
sample indicated large ${\cal S}_{\pi\pi}$ 
and ${\cal A}_{\pi\pi}$ values~\cite{ref:belle-140-pipi},
while no significant $CP$ asymmetry was observed by the BaBar
Collaboration~\cite{ref:babar-pipi}.
It is therefore important to measure the $CP$-violating parameters
with larger statistics. 

The measurement in this Letter is based on a $253~$fb$^{-1}$ data sample
containing $275 \times 10^6$~$B\overline{B}$ pairs collected with the
Belle detector at the KEKB $e^+e^-$ asymmetric-energy (3.5 on 8~GeV)
collider~\cite{ref:kekb} operating at the $\Upsilon(4S)$ resonance.
The $\Upsilon(4S)$ is produced with a Lorentz boost factor of
$\beta\gamma = 0.425$ along the $z$ axis, which is antiparallel
to the positron beam direction.
Since the two $B$ mesons are produced nearly at rest in the 
$\Upsilon(4S)$ center-of-mass system (CMS),
the decay time difference $\Delta t$ is determined from the distance
between the two $B$ meson decay positions along the 
$z$-direction ($\Delta z$):
$\Delta t \cong \Delta z / c\beta\gamma$, where $c$ is the velocity of light.

The Belle detector~\cite{ref:belle-detector}
is a large-solid-angle magnetic spectrometer 
that consists of a silicon vertex detector (SVD),
a 50-layer central drift chamber (CDC),
an array of aerogel threshold Cherenkov counters (ACC),
a barrel-like arrangement of time-of-flight scintillation 
counters (TOF),
and an electromagnetic calorimeter (ECL)
comprised of CsI(Tl) crystals located inside a superconducting
solenoid coil that provides a 1.5~T magnetic field.
An iron flux-return located outside of the coil is 
instrumented to detect $K^0_L$ mesons and to identify muons (KLM).
A sample containing $152 \times 10^6$ $B\overline{B}$ pairs (Set I)
was collected with a 2.0~cm radius beampipe and a 3-layer
silicon vertex detector, while
a sample with $123 \times 10^6$ $B\overline{B}$ pairs (Set II)
was collected with a 1.5~cm radius beampipe, 
a 4-layer silicon detector, and
a small-cell inner drift chamber~\cite{ref:svd2}.

We reconstruct $B^0 \to \pi^+\pi^-$ candidates using oppositely
charged track pairs that are positively identified as pions
by combining information from the ACC and the CDC $dE/dx$
measurements.
The pion detection efficiency is 90\%, and 11\% of kaons are
misidentified as pions.
We select $B$ meson candidates using the energy difference 
$\Delta E \equiv E_B^*-E_{\rm beam}^*$
and the beam-energy constrained mass 
$M_{\rm bc}\equiv\sqrt{(E_{\rm beam}^*)^2-(p_B^*)^2}$,
where $E_{\rm beam}^*$ is the CMS beam-energy, and 
$E_B^*$ and $p_B^*$ are the CMS energy and momentum of the $B$ candidate.
We define the  signal region as
$5.271$~GeV/$c^2 < M_{\rm bc} < 5.287$~GeV/$c^2$ and
$|\Delta E|<0.064$~GeV, which corresponds to $\pm3$ 
standard deviations ($\sigma$) from the central values.

We identify the flavor of the accompanying $B$ meson 
from inclusive properties of particles that are not
associated with the reconstructed $B^0 \to \pi^+\pi^-$ decay.
We use $q$ defined in Eq.~(\ref{eq:signal_pdf}) and $r$ to
represent the tagging information.
The parameter $r$ is an event-by-event, Monte Carlo (MC) determined
flavor-tagging dilution factor that ranges from $r=0$ for no flavor
discrimination to $r=1$ for unambiguous flavor assignment.
It is used only to sort data into six $r$ intervals.
The wrong tag fractions for the six $r$ intervals, 
$w_l$ $(l=1,6)$, and the
differences between $B^0$ and $\overline{B}{}^0$ decays,
$\Delta w_l$, are determined from 
data~\cite{ref:flavor-tagging, ref:btos2004}.

To suppress the continuum background ($e^+e^- \to q\overline{q};
q = u,d,s,c$), 
we apply the technique used in Ref.~\cite{ref:belle-140-pipi}.
We form a likelihood ratio $LR$ based on event
topology variables and impose requirements on $LR$
to suppress continuum events. The $LR$ requirement is
determined by optimizing the expected sensitivity
using MC signal events and events in the sideband region
in
$5.20~{\rm GeV}/c^2<M_{\rm bc}<5.26~{\rm GeV}/c^2$
or $+0.1~{\rm GeV}<\Delta E<+0.5~{\rm GeV}$.
As the tag-dilution variable $r$ also suppresses continuum
events, we optimize $LR$ separately for each of the
$r$ bins. For all $r$ bins we accept events having
$LR>0.86$; we then include additional events having
lower $LR$ thresholds of 0.50, 0.45, 0.45, 0.45, 0.45,
and 0.20, respectively, for the six $r$ bins. There are
thus 12 distinct bins of $LR$-$r$ for selected events.

\begin{figure}
\begin{center}
\resizebox{0.46\columnwidth}{!}{\includegraphics{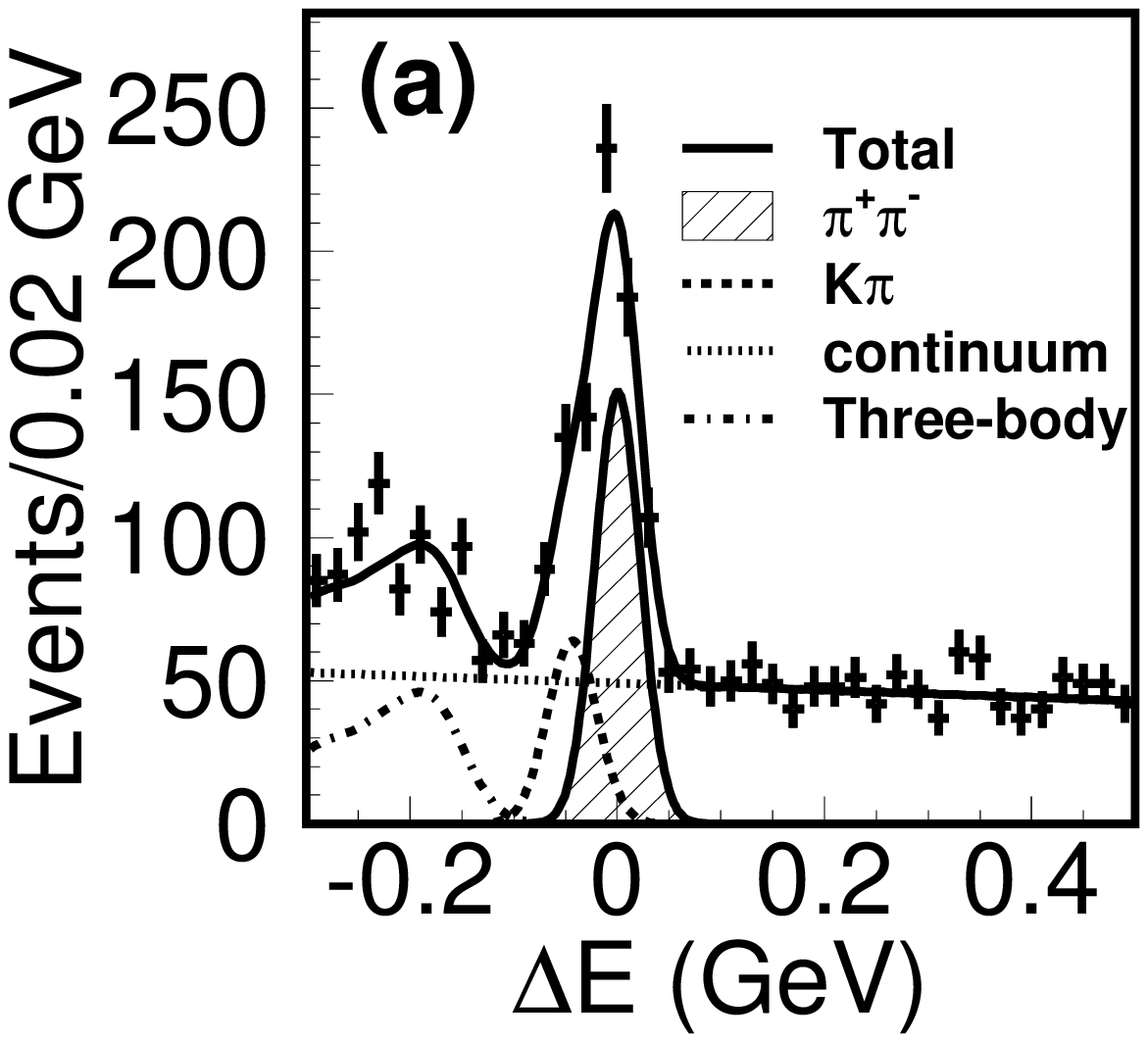}}
\resizebox{0.46\columnwidth}{!}{\includegraphics{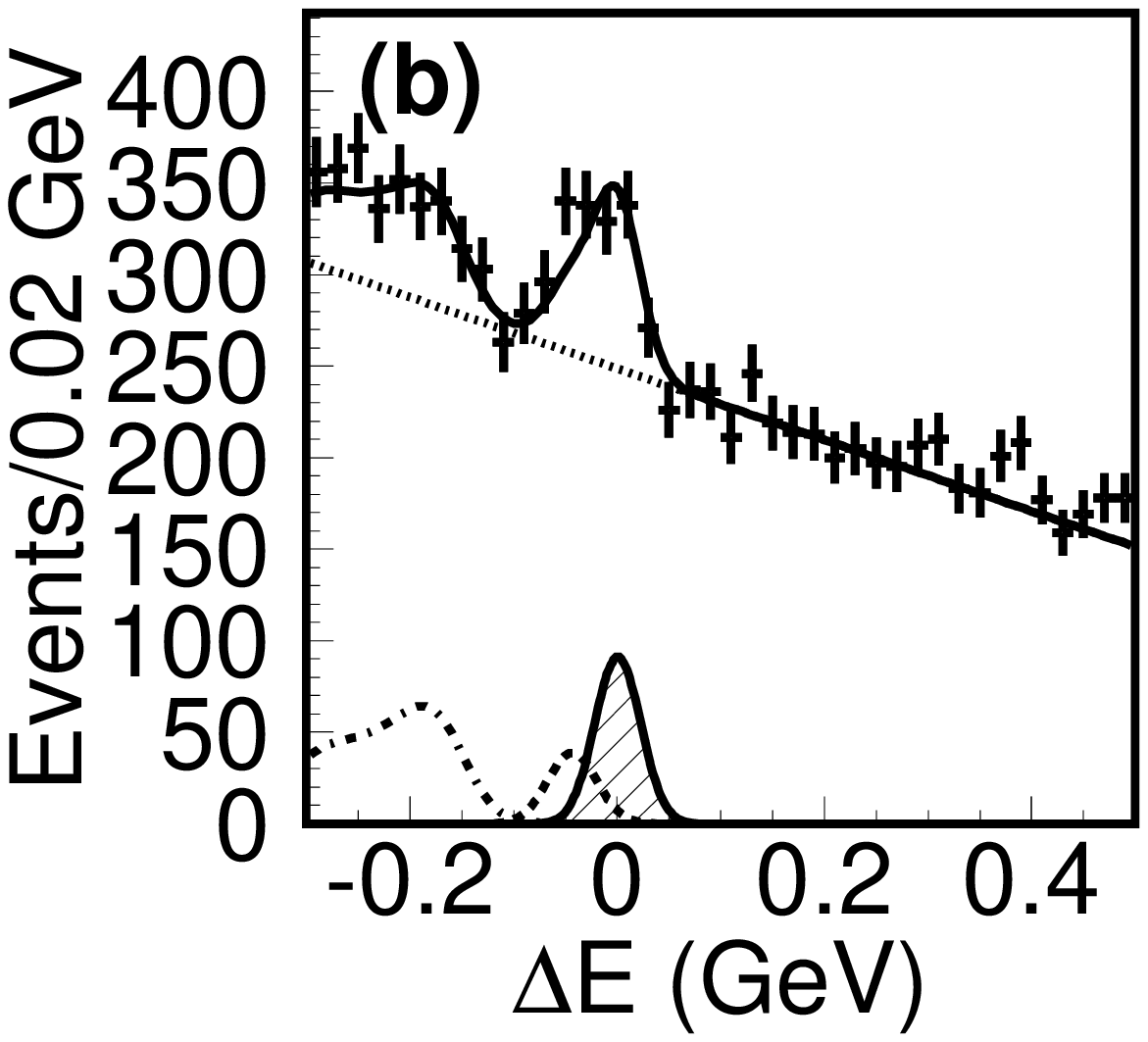}}
\caption{ $\Delta E$ distributions in the $M_{\rm bc}$
signal region for $B^0\to\pi^+\pi^-$ candidates with (a) $LR>0.86$
and (b) $LR<0.86$.
}
\label{fig:delta-e}
\end{center}
\end{figure}

We extract 2,820 signal candidates
by applying the above requirements and the vertex reconstruction algorithm
used in Ref.~\cite{ref:btos2004} to the data sample.
Figure~\ref{fig:delta-e} shows the $\Delta E$ distributions
for the events with (a) $LR>0.86$ and (b) $LR<0.86$ 
in the $M_{\rm bc}$ signal region.
The $B^0\to\pi^+\pi^-$ signal yield is determined from
an unbinned two-dimensional maximum likelihood fit to the
$\Delta E$-$M_{\rm bc}$ distribution in the range of
$M_{\rm bc}>5.20~{\rm GeV}/c^2$ and
$-0.3~{\rm GeV}<\Delta E<+0.5~{\rm GeV}$ with 
signal events plus contributions from misidentified
$B^0\to K^+\pi^-$ events, 
the continuum background, and three-body $B$ decays.
We use a single Gaussian for the signal and $B^0\to K^+\pi^-$
events in $\Delta E$ and $M_{\rm bc}$.
The continuum background shapes in $\Delta E$ and $M_{\rm bc}$ 
are described by 
a first-order polynomial  and an ARGUS function~\cite{ref:argus},
respectively.
For the three-body $B$ decay background shape,
we employ a smoothed two-dimensional histogram obtained from a
large MC sample.
The fit to the subset with $LR>0.86$ yields $415\pm27$ $\pi^+\pi^-$ events and
$154\pm19$ $K^+\pi^-$ events in the signal region, where
the errors are statistical only.
The $K^+\pi^-$ contamination is consistent with the $K\to\pi$
misidentification probability, which is measured independently.
Extrapolating from the size of the continuum background in this fit,
we expect $315\pm3$ continuum events in the signal region.
We use MC-determined fractions as in~\cite{ref:belle-140-pipi}
to calculate the numbers of decays for $LR<0.86$.
We expect $251\pm16$ $\pi^+\pi^-$, $93\pm12$ $K^+\pi^-$ 
and $1,592\pm15$ continuum events in the signal region.
The contribution from three-body $B$ decays is negligibly small
in the signal region.

We determine ${\cal S}_{\pi\pi}$ and ${\cal A}_{\pi\pi}$
by applying an unbinned maximum likelihood fit to the distribution of
proper-time difference $\Delta t$.
The probability density function (PDF) for the signal events
is given in Eq.~(\ref{eq:signal_pdf})
modified to incorporate the effect 
of incorrect flavor assignment $w_l$ and $\Delta w_l$.
The distribution is convolved with the proper time interval
resolution function $R_{\rm sig}(\Delta t)$ in order to take into
account the finite 
position resolution ~\cite{ref:resolution-function, ref:btos2004}.
The PDF for $B^0\to K^+\pi^-$ is 
${\cal P}_{K\pi}^q(\Delta t, w_l, \Delta w_l)
=(1/4\tau_{B^0})e^{-|\Delta t|/\tau_{B^0}}[1-q\Delta w_l + q(1-2w_l)
{\cal A}_{K\pi}^{\rm eff}\cos(\Delta m_d\Delta t)]$.
We use ${\cal A}_{K\pi}^{\rm eff}=
({\cal A}_{K\pi}+{\cal A}_{\varepsilon})/
(1+{\cal A}_{K\pi}{\cal A}_{\varepsilon})$,
where ${\cal A}_{K\pi}=-0.109\pm0.019$ 
is the measured direct $CP$-violating parameter in $B^0\to K^+\pi^-$ decays
~\cite{ref:hfag}, and ${\cal A}_{\varepsilon}$ is the difference in
the product of the pion efficiency and 
kaon misidentification probability between
$\pi^+(K^-)$ and $\pi^-(K^+)$ divided by their sum ~\cite{ref:kpi-acp}.
The inclusion of ${\cal A}_{\varepsilon}$ changes the ${\cal A}_{K\pi}$ value
by 11\%.
We make use of the same resolution function $R_{\rm sig}(\Delta t)$
for the $B^0\to K^+\pi^-$ events.
The PDF for the continuum background events is
${\cal P}_{q\overline{q}}(\Delta t)
= 1/2\cdot(1+q{\cal A}_{q\overline{q}})
[(f_{\tau}/2\tau_{q\overline{q}})e^{-|\Delta t|/\tau_{q\overline{q}}}
+(1-f_{\tau})\delta(\Delta t)]$,
where $f_{\tau}$ is the fraction of the background with effective
lifetime $\tau_{q\overline{q}}$, and $\delta$ is the Dirac delta function.
We use ${\cal A}_{q\overline{q}}=0$ as a default.
A fit to the sideband events 
yields
${\cal A}_{q\overline{q}}=+0.01\pm0.01$ $(-0.00\pm0.01)$
for the data in Set I (II).
This uncertainty in the background asymmetry is included in the systematic error
for the ${\cal S}_{\pi\pi}$ and ${\cal A}_{\pi\pi}$ measurement.
The background PDF ${\cal P}_{q\overline{q}}$ is convolved with
a background resolution function $R_{q\overline{q}}$.
All parameters in ${\cal P}_{q\overline{q}}$ and $R_{q\overline{q}}$
are determined from sideband events.

We define a likelihood value for each ($i$th) event as a
function of ${\cal S}_{\pi\pi}$ and ${\cal A}_{\pi\pi}$:
\begin{eqnarray}
%
%
P_i &=& (1-f_{\rm ol})\int_{-\infty}^{+\infty}
[\{
  f^m_{\pi\pi}{\cal P}_{\pi\pi}^q(\Delta t^{\prime}, w_l, \Delta w_l; 
                {\cal S}_{\pi\pi}, {\cal A}_{\pi\pi}) \nonumber \\
 & & +f^m_{K\pi}{\cal P}_{K\pi}^q(\Delta t^{\prime}, w_l, \Delta w_l)
 \}R_{\rm sig}(\Delta t_i - \Delta t^{\prime})\nonumber \\
 & & +f_{q\overline{q}}^m{\cal P}_{q\overline{q}}(\Delta t^{\prime})
  R_{q\overline{q}}(\Delta t_i - \Delta t^{\prime})
]d\Delta t^{\prime}\nonumber \\
& & +f_{\rm ol}{\cal P}_{\rm ol}(\Delta t_i).
%
%
\end{eqnarray}
Here, the probability functions 
$f^m_k(k=\pi\pi, K\pi,{\rm or~} q\overline{q})$
are determined on an event-by-event basis as functions of
$\Delta E$ and $M_{\rm bc}$ for each $LR$-$r$ bin ($m=1,12$).
A small number of signal and background events that have
large values of $\Delta t$ is accommodated by the outlier PDF,
${\cal P}_{\rm ol}$, with a fractional area $f_{\rm ol}$.
In the fit, ${\cal S}_{\pi\pi}$ and ${\cal A}_{\pi\pi}$
are the only free parameters and are determined by maximizing the likelihood
function ${\cal L}=\prod_i P_i$, where the product is over all the
$B^0\to\pi^+\pi^-$ candidates.

\begin{figure}
\begin{center}
\resizebox{0.6\columnwidth}{!}{\includegraphics{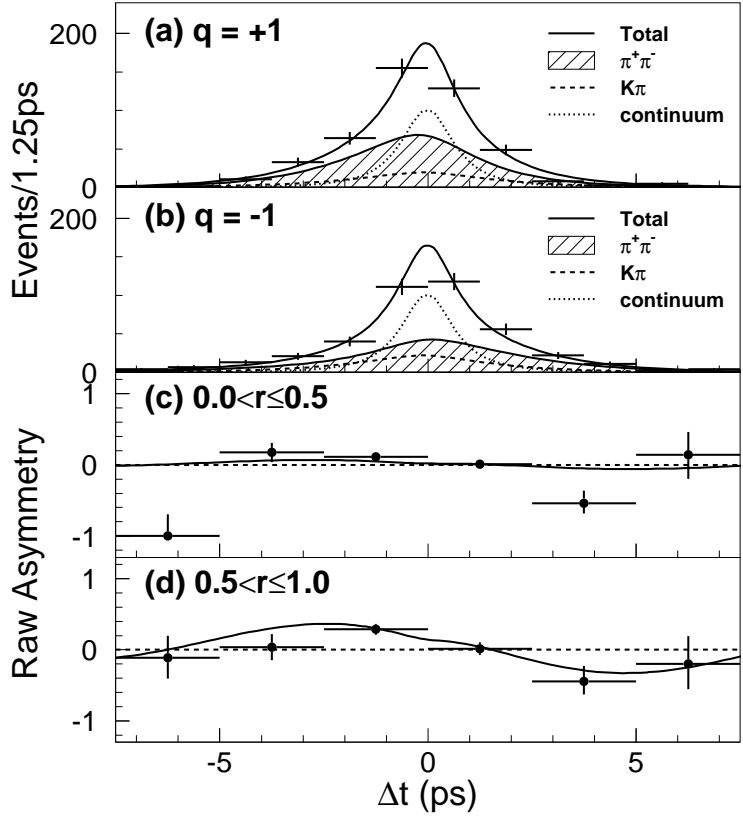}}
\caption{ $\Delta t$ distributions for the 884
$B^0\to \pi^+\pi^-$ candidates with $LR>0.86$ in the signal region:
(a) 470 candidates with $q=+1$, 
(b) 414 candidates with $q=-1$.
Raw asymmetry, ${\cal A}_{CP}$, in each $\Delta t$ bin with
(c) $0 < r \leq 0.5$ and (d) $0.5 < r \leq 1.0$.
The solid lines show  the results of the unbinned maximum likelihood
fit to the $\Delta t$ distribution of the 2,820 $B^0\to \pi^+\pi^-$
candidates.}
\label{fig:cp-asymmetry}
\end{center}
\end{figure}

The unbinned maximum likelihood fit to the 2,820 $B^0\to\pi^+\pi^-$
candidates 
containing $666\pm43$ $\pi^+\pi^-$ signal events
(1,486 $B^0$ tags and 1,334 $\overline{B}{}^0$ tags)
yields 
${\cal S}_{\pi\pi}=\svalue\sstaterr({\rm stat})\ssysterr({\rm syst})$ and
${\cal A}_{\pi\pi}=\avalue\astaterr({\rm stat})\asysterr({\rm syst})$.
The correlation between ${\cal S}_{\pi\pi}$ and ${\cal A}_{\pi\pi}$
is $\ascorrelation$.
In this Letter, we quote the usual fit errors from the likelihood functions,
called the MINOS errors, as statistical uncertainties~\cite{footnote:MINOS}.
Figures~\ref{fig:cp-asymmetry}(a) and \ref{fig:cp-asymmetry}(b) 
show the $\Delta t$ distributions
for the 470 $B^0$- and 414 $\overline{B}{}^0$-tagged events in the
subset of data with $LR>0.86$.
We define the raw asymmetry ${\cal A}_{CP}$ in each $\Delta t$ bin
by ${\cal A}_{CP}=(N_+-N_-)/(N_++N_-)$, where
$N_{+(-)}$ is the number of observed candidates with $q=+1$ $(-1)$.
Figures~\ref{fig:cp-asymmetry}(c) and \ref{fig:cp-asymmetry}(d) 
show the raw asymmetries for two regions of the flavor-tagging
parameter $r$.

The main contributions to the systematic error 
are due to the uncertainties in the
vertex reconstruction ($\pm0.04$ for ${\cal S}_{\pi\pi}$ and
$^{+0.03}_{-0.01}$ for ${\cal A}_{\pi\pi}$) and
event fraction ($\pm0.02$ for ${\cal S}_{\pi\pi}$ and
$\pm0.04$ for ${\cal A}_{\pi\pi}$); the latter includes
the uncertainties in ${\cal A}_{q\overline{q}}$ and
final state radiation.
We include the effect of tag side interference~\cite{ref:tag-side-int}
on ${\cal S}_{\pi\pi}(\pm0.01)$ and
${\cal A}_{\pi\pi}(^{+0.02}_{-0.04})$.
Other sources of systematic error are the uncertainties
in the wrong tag fraction 
($\pm0.01$ for ${\cal S}_{\pi\pi}$ and $\pm0.01$ for ${\cal A}_{\pi\pi}$), 
physics parameters ($\tau_{B^0}$,
$\Delta m_d$ and ${\cal A}_{K\pi}$)
($<0.01$ for ${\cal S}_{\pi\pi}$ and $\pm0.01$ for ${\cal A}_{\pi\pi}$), 
resolution function
($\pm0.04$ for ${\cal S}_{\pi\pi}$ and $\pm0.01$ for ${\cal A}_{\pi\pi}$), 
background $\Delta t$ shape
($<0.01$ for ${\cal S}_{\pi\pi}$ and $<0.01$ for ${\cal A}_{\pi\pi}$), 
and fit bias
($\pm0.01$ for ${\cal S}_{\pi\pi}$ and $\pm0.01$ for ${\cal A}_{\pi\pi}$).
We add each contribution in quadrature to obtain the total
systematic error.

We carry out a number of checks to validate our results.
The $B^0$ lifetime is measured with $B^0\to \pi^+\pi^-$ candidates.
The result is $\tau_{B^0}=1.50\pm0.07$~ps, consistent with the
world average value~\cite{ref:PDBook2004}.
The $CP$ fit to the sideband events yields
no significant asymmetry.
We check the measurement of ${\cal A}_{\pi\pi}$ using
a time-integrated fit and
obtain ${\cal A}_{\pi\pi}=+0.52\pm0.14$, consistent
with the time-dependent fit result.
We also select $B^0\to K^+\pi^-$ candidate events with
charged tracks positively identified as kaons that have
a topology similar to the $B^0\to \pi^+\pi^-$ signal events.
The $CP$ fit to the 4,293 $B^0\to K^+\pi^-$ 
candidates (2,207 signal events) yields
${\cal S}_{K\pi}=+0.09\pm0.08$, consistent with zero, and
${\cal A}_{K\pi}=-0.06\pm0.06$, in agreement with the world
average value~\cite{ref:hfag}.
With the $K^+\pi^-$ sample, we determine
$\tau_{B^0}=1.51\pm0.04$~ps and $\Delta m_d=0.46\pm0.03$~ps$^{-1}$,
which are also in agreement with the world average values~\cite{ref:PDBook2004}.

To determine the statistical significance of our measurement,
we apply the frequentist procedure described in
Ref.~\cite{ref:belle-140-pipi} that takes into account 
both statistical and systematic errors.
The hypothesis of $CP$ symmetry conservation,
${\cal S}_{\pi\pi}={\cal A}_{\pi\pi}=0$, is ruled out at
a confidence level (C.L.) of
$1-$C.L.$=5.6\times 10^{-8}$,
equivalent to 
a $5.4\sigma$ significance for one-dimensional Gaussian errors.
The case of no direct $CP$ violation, ${\cal A}_{\pi\pi}=0$,
is also ruled out with a significance greater than
$4.0\sigma$ for any ${\cal S}_{\pi\pi}$ value.

\begin{figure}
\begin{center}
\resizebox{0.46\columnwidth}{!}{\includegraphics{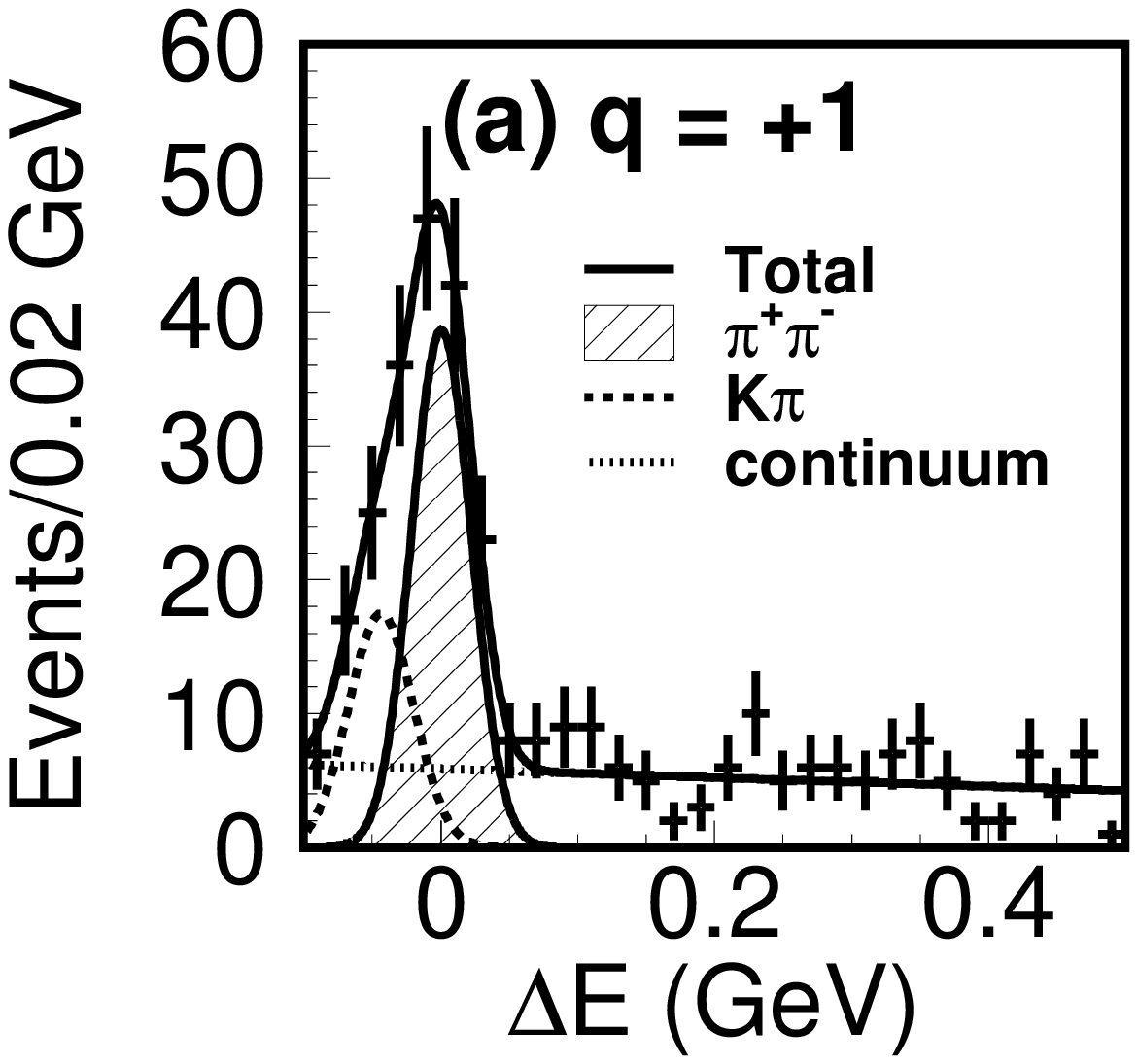}}
\resizebox{0.46\columnwidth}{!}{\includegraphics{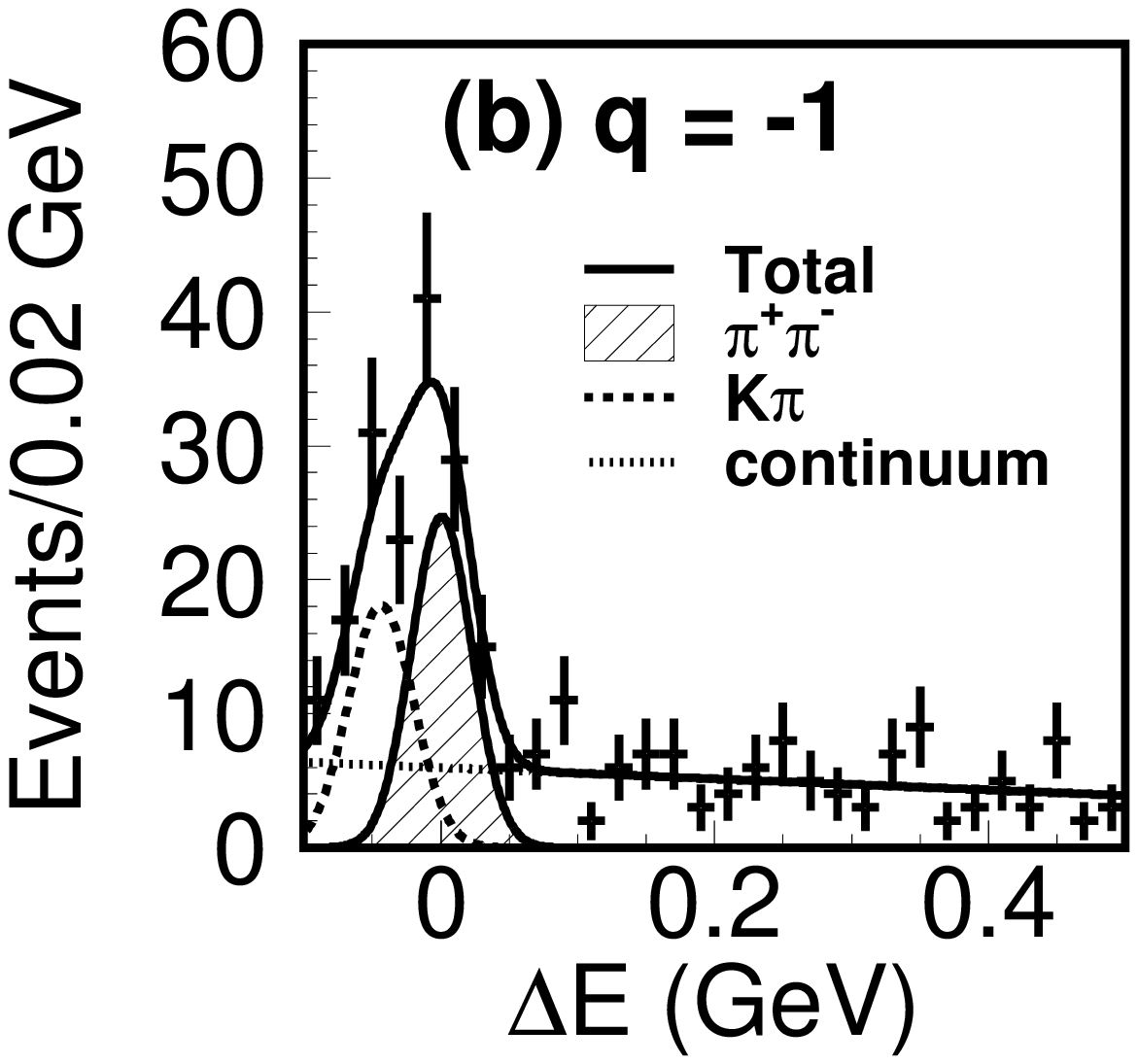}}
\caption{
$\Delta E$ distributions in the $M_{\rm bc}$ signal region
for the $B^0\to \pi^+\pi^-$ candidates
with $LR>0.86$ and $0.5< r \leq 1.0$ for (a) $q=+1$ and (b) $q=-1$.
}
\label{fig:de-direct-cp}
\end{center}
\end{figure}

Figure~\ref{fig:de-direct-cp} shows the $\Delta E$ distributions
for $B^0\to \pi^+\pi^-$ candidates with $LR>0.86$ and $0.5< r \leq 1.0$
for (a) $q=+1$ and (b) $q=-1$ subsets in the $M_{\rm bc}$ signal region.
An unbinned two-dimensional maximum likelihood fit 
to the $q=+1$ ($q=-1$) subset
yields $107\pm13$ ($69\pm11$) $\pi^+\pi^-$,
$42\pm9$ ($43\pm9$) $K^+\pi^-$ and
$38\pm1$ ($38\pm1$) continuum events in the signal box.
The $K^+\pi^-$ and continuum background yields are consistent
between the two subsets as expected, while the $\pi^+\pi^-$ yields
are significantly different; thus
large direct $CP$-violation in $B^0\to \pi^+\pi^-$ decays is 
manifest in the contrast of the two subsets.
These results also support the expectation from
$SU(3)$ symmetry that 
${\cal A}_{\pi\pi} \sim -3 \cdot
{\cal A}_{K\pi}$~\cite{ref:gronau-rosner-2004}.
%
%

We constrain the ratio of the magnitude of the penguin to 
tree amplitudes $|P/T|$
and the strong phase difference $\delta\equiv \delta_P-\delta_T$ 
by adopting the notation of Ref.~\cite{ref:pt-relation},
where $\delta_{P(T)}$ is the strong phase of the penguin (tree) amplitude.
By using $\phi_1= 23.5 \pm 1.6^{\circ}$~\cite{ref:hfag}, 
we find 95.4\% confidence intervals of
$|P/T|>0.17$ and $-180^{\circ}<\delta<-4^{\circ}$.

\begin{figure}
\begin{center}
\resizebox{0.6\columnwidth}{!}{\includegraphics{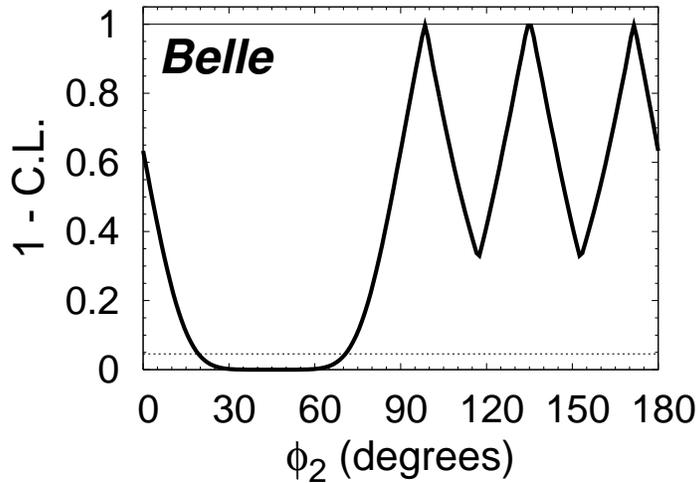}}
\caption{Confidence level as a function of the CKM angle $\phi_2$
obtained with an isospin analysis
using Belle measurements of ${\cal S}_{\pi\pi}$
and ${\cal A}_{\pi\pi}$.
The dotted line indicates C.L. = 95.4\%.
}
\label{fig:phi2}
\end{center}
\end{figure}

To constrain $\phi_2$, we employ isospin 
relations~\cite{ref:isospin} and 
the approach of Ref.~\cite{ref:ckmfit} for the statistical treatment.
We use the measured branching ratios 
of $B^0\to \pi^+\pi^-$, $\pi^0\pi^0$ and $B^+\to \pi^+\pi^0$, and
the direct $CP$-asymmetry for $B^0\to \pi^0\pi^0$~\cite{ref:hfag}
as well as our measured values of ${\cal S}_{\pi\pi}$ 
and ${\cal A}_{\pi\pi}$ taking into account their correlation.
Figure~\ref{fig:phi2} shows the obtained C.L. as a function of $\phi_2$.
We find an allowed range for $\phi_2$ at 95.4\% C.L. of
$0^{\circ}<\phi_2<19^{\circ}$ and $71^{\circ}<\phi_2<180^{\circ}$.

In summary, we have performed a new measurement of
the $CP$-violating parameters in $B^0\to \pi^+\pi^-$ decays
using a $253$~fb$^{-1}$ data sample.
We obtain 
${\cal S}_{\pi\pi}=\svalue\sstaterr({\rm stat})\ssysterr({\rm syst})$ and
${\cal A}_{\pi\pi}=\avalue\astaterr({\rm stat})\asysterr({\rm syst})$.
We rule out the $CP$-conserving case, 
${\cal S}_{\pi\pi}={\cal A}_{\pi\pi}=0$, at the $5.4\sigma$ level.
We find compelling evidence for direct $CP$ asymmetry with
$4.0\sigma$ significance.
The results confirm the previous Belle measurement of the $CP$-violating 
parameters
as well as the earlier evidence for direct $CP$ violation
in $B^0\to \pi^+\pi^-$ decays~\cite{ref:belle-140-pipi}.

We thank the KEKB group for the excellent operation of the
accelerator, the KEK cryogenics group for the efficient
operation of the solenoid, and the KEK computer group and
the NII for valuable computing and Super-SINET network
support.  We acknowledge support from MEXT and JSPS (Japan);
ARC and DEST (Australia); NSFC (contract No.~10175071,
China); DST (India); the BK21 program of MOEHRD and the CHEP
SRC program of KOSEF (Korea); KBN (contract No.~2P03B 01324,
Poland); MIST (Russia); MESS (Slovenia);  SNSF (Switzerland); NSC and MOE
(Taiwan); and DOE (USA).


\clearpage
\newpage


\begin{thebibliography}{999}

\bibitem{ref:km}
M.~Kobayashi and T.~Maskawa, Prog. Theor. Phys. {\bf 49}, 652 (1973).

\bibitem{ref:tcpv}
A.B.~Carter and A.I.~Sanda, Phys. Rev. D {\bf 23}, 1567 (1981); 
I.I.~Bigi and A.I.~Sanda, Nucl. Phys. B {\bf 193}, 85 (1981).

\bibitem{ref:charge-conjugate}
Throughout this Letter, the inclusion of the charge conjugate
decay mode is implied unless otherwise stated.

\bibitem{ref:belle-140-pipi}
Belle Collaboration, K.~Abe {\it et al.}, 
Phys. Rev. Lett. {\bf 93}, 021601 (2004);
see also Belle Collaboration, K.~Abe {\it et al.}, 
Phys. Rev. D {\bf 68}, 012001 (2003) for results based on
a 78 fb$^{-1}$ data sample.
The results reported here supersede
those of these two publications.

\bibitem{ref:babar-pipi}
BaBar Collaboration, B.~Aubert {\it et al.}, 
hep-ex/0501071, submitted to Phys. Rev. Lett.

\bibitem{ref:kekb}
S.~Kurokawa and E.~Kikutani {\it et al.}, Nucl. Instr. and Meth. A
{\bf 499}, 1 (2003).

\bibitem{ref:belle-detector}
Belle Collaboration, A.~Abashian {\it et al.}, Nucl. Instr. and Meth. A
{\bf 479}, 117 (2002).

\bibitem{ref:svd2}
Y.~Ushiroda (Belle SVD2 Group), Nucl. Instr. and Meth. A {\bf 511}, 6 (2003).

\bibitem{ref:flavor-tagging}
H.~Kakuno {\it et al.}, Nucl. Instr. and Meth. A {\bf 533}, 516 (2004).

\bibitem{ref:btos2004}
Belle Collaboration, K.-F.~Chen {\it et al.}, hep-ex/0504023,
submitted to Phys. Rev. D.

\bibitem{ref:argus}
ARGUS Collaboration, H.~Albrecht {\it et al.},
Phys. Lett. B {\bf 241}, 278 (1990).

\bibitem{ref:resolution-function}
H.~Tajima {\it et al.}, Nucl. Instr. and Meth. A {\bf 533}, 370 (2004).

\bibitem{ref:hfag}
The Heavy Flavor Averaging Group (HFAG), hep-ex/0412073,
http://www.slac.stanford.edu/xorg/hfag (2004).

\bibitem{ref:kpi-acp}
Belle Collaboration, Y.~Chao {\it et al.}, 
Phys. Rev. Lett. {\bf 93}, 191802 (2004).

\bibitem{footnote:MINOS}
The rms values of the ${\cal S}_{\pi\pi}$
and ${\cal A}_{\pi\pi}$ distributions of MC pseudo-experiments
were quoted as the statistical uncertainties in the previous
publications~\cite{ref:belle-140-pipi}. 
With improved statistics, we find that MINOS errors are
approximately symmetric and agree well with the rms values
(0.15 for ${\cal S}_{\pi\pi}$ and 0.11 for ${\cal A}_{\pi\pi}$).

\bibitem{ref:tag-side-int}
O.~Long, M.~Baak, R.N.~Cahn, and D.~Kirkby, 
Phys. Rev. D {\bf 68}, 034010 (2003).

\bibitem{ref:PDBook2004}
Particle Data Group (S.~Eidelman {\it et al.}), 
Phys. Lett. B {\bf 592}, 1 (2004).

\bibitem{ref:gronau-rosner-2004}
M.~Gronau and J.L.~Rosner,
Phys. Lett. B {\bf 595}, 339 (2004),
N.G.~Deshpande and X.G.~He, 
Phys. Rev. Lett. {\bf 75}, 1703 (1995).

\bibitem{ref:pt-relation}
M.~Gronau and J.L.~Rosner, Phys. Rev. D {\bf 65}, 093012 (2002).

\bibitem{ref:isospin}
M.~Gronau and D.~London, Phys. Rev. Lett. {\bf 65}, 3381 (1990).

\bibitem{ref:ckmfit}
J.~Charles {\it et al.}, hep-ph/0406184,
accepted by Eur. Phys. J. {\bf C}.

\end{thebibliography}
\end{document}